\documentclass[12pt]{reportj}
\usepackage{deluxetablej}
\usepackage{hyperref} %% Links all your references to pages
\makeatletter
\def\@to{to}
\makeatother
\usepackage{times}
\usepackage{graphicx}
\usepackage{xcolor}
\usepackage{tocloft}
\usepackage{fancyheadings}
\usepackage{tablefootnote}
\usepackage{adjustbox}

\emergencystretch=\maxdimen
\usepackage{datetime}
\newdateformat{ddmonthyyyy}{\THEDAY\ \monthname[\THEMONTH]\ \THEYEAR}

\renewcommand\thesection{\arabic{section}}
\renewcommand\thesubsection{\thesection.\arabic{subsection}}

\def\ssection#1{\setcounter{subsection}{0} \refstepcounter{section} \section*{\hbox to \hsize{\large\bf \arabic{section}. #1\hfill }}\label{sec} \addcontentsline{toc}{section}{\arabic{section}. #1}}
\def\ssubsection#1{\setcounter{subsubsection}{0} \refstepcounter{subsection}\subsection*{\hbox to \hsize{\normalsize\bfseries\itshape \arabic{section}.\arabic{subsection} #1\hfill}}\label{subsec} \addcontentsline{toc}{subsection}{\arabic{section}.\arabic{subsection} #1}}
\def\ssubsubsection#1{\refstepcounter{subsubsection}\subsection*{\hbox to \hsize{\normalsize\it \arabic{section}.\arabic{subsection}.\arabic{subsubsection} #1\hfill}}\label{subsubsec} \addcontentsline{toc}{subsubsection}{\arabic{section}.\arabic{subsection}.\arabic{subsubsection} #1}}

\def\ssectionstar#1{\section*{\hbox to \hsize{\large\bf #1\hfill}} \addcontentsline{toc}{section}{#1}}
\def\ssubsectionstar#1{\subsection*{\hbox to \hsize{\normalsize\bfseries\itshape #1\hfill}} \addcontentsline{toc}{subsection}{#1}}
\def\ssubsubsectionstar#1{\subsection*{\hbox to \hsize{\normalsize\it  #1\hfill}} \addcontentsline{toc}{subsection}{#1}}

\renewcommand{\cftaftertoctitle}{%
\mbox{}\hfill{\normalfont Page}}
\begin{document}

~\\

% ST Logo in the top left
\vspace{-2.4cm}
\noindent\includegraphics*[width=0.295\linewidth]{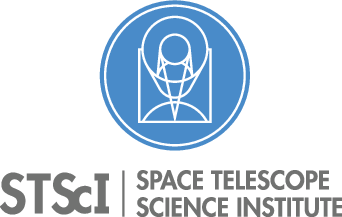}

\vspace{-0.4cm}

\begin{flushright}
    %%% Put the instrument, year, and ISR number here (and also below) %%%
    {\bf Instrument Science Report STIS 2024-02}
    
    \vspace{1.1cm}
    
    %%% Put ISR Title %%%
    {\bf\Huge Recalibration of Pre-SM4 STIS Echelle Throughputs}
    
    \rule{0.25\linewidth}{0.5pt}
    
    \vspace{0.5cm}
    
    %Put Authors
    Matthew R. Siebert $^1$,  Joleen K. Carlberg $^1$, Svea Hernandez $^1$, and TalaWanda Monroe \rlap{$^1$}
    \linebreak
    \newline
    %Put Author's affiliations
    \footnotesize{$^1$ Space Telescope Science Institute, Baltimore, MD\\
                        }
    
    \vspace{0.5cm}
    
    % Date in DD Month YYYY format based on when you compile it
    15 March 2024
     %\ddmonthyyyy\today 
\end{flushright}

\vspace{0.1cm}

%%% ABSTRACT %%%
\noindent\rule{\linewidth}{1.0pt}
\noindent{\bf A{\footnotesize BSTRACT}}

{\it \noindent Recent improvements to stellar atmospheric models have merited updated flux calibration for high priority STIS observing modes. Specifically, in the FUV and NUV, continuum differences of 1-3\% are present between the newest models (CALSPECv11) and previous models (CALSPECv04-v07). As a result of these improvements the STIS team has derived updated sensitivity curves and blaze shift coefficients for a variety of echelle modes in order to meet targeted flux accuracies. The first series of echelle sensitivity updates primarily targeted post-Servicing Mission 4 (SM4; in 2009) observations. In this ISR, we investigate instead applying a simple scaling (derived from the ratio of new vs old CALSPEC model continua) to the previously determined throughputs of STIS echelle modes. This alternative approach has a straightforward implementation and provides reasonable accuracy, especially in cases where available calibration data are lacking (e.g., pre-SM4 era). Adopting this scaling approach, we delivered pre-SM4 throughput updates for 8 echelle modes, resulting in typical improvements of 0.5-2.4\% across the FUV and NUV. }

\vspace{-0.1cm}
\noindent\rule{\linewidth}{1.0pt}

%% Table of Contents
%% Need to compile twice to get the page numbers correct
\renewcommand{\cftaftertoctitle}{\thispagestyle{fancy}}
\tableofcontents

%%% MAIN TEXT BELOW %%%

\vspace{-0.3cm}
\ssection{Introduction}\label{sec:Introduction}

The Space Telescope Imaging Spectrograph is a versatile instrument capable of observing at FUV and NUV wavelengths with 44 different echelle settings at medium to high resolution. The STIS instrument failed in 2004, and was brought back to operations during Servicing Mission 4 (SM4) in May 2009. A special calibration program was implemented (Cycle 17, 11866) to observe with all echelle settings to derive new sensitivities for post-SM4 data. This calibration was performed with the atmospheric model of the white dwarf standard star for G191-B2B in the CALSPECv07 library. The details of this update are in \href{https://www.stsci.edu/files/live/sites/www/files/home/hst/instrumentation/stis/documentation/instrument-science-reports/_documents/2012_01.pdf}{Bostroem et al. 2012}. 

Since then, the stellar atmospheric models for the three primary standard stars (GD 71, GD 153, G 191-B2B) have been updated \href{https://ui.adsabs.harvard.edu/abs/2020AJ....160...21B/abstract}{(Bohlin et al.\ 2020)}, with the latest improvements included in the CALSPECv11 database\footnote{\href{https://www.stsci.edu/hst/instrumentation/reference-data-for-calibration-and-tools/astronomical-catalogs/calspec}{https://www.stsci.edu/hst/instrumentation/reference-data-for-calibration-and-tools/astronomical-catalogs/calspec}}. These updates changed the continuum flux of the standard stars by as much as $3\%$, which necessitated the recalibration of all STIS observing modes. The STIS team has prioritized the most used echelle modes and has updated sensitivity curves and blaze shift coefficients for E140M/1425 (see \href{https://www.stsci.edu/contents/news/stis-stans/april-2022-stan#article1}{April 2022 STAN}, \href{https://www.stsci.edu/files/live/sites/www/files/home/hst/instrumentation/stis/documentation/instrument-science-reports/_documents/2022-04.pdf}{Carlberg et al. 2022}); E230M/1978, E230M/2707, E230M/2415 (see \href{https://www.stsci.edu/contents/news/stis-stans/january-2023-stan#article1}{Jan 2023 STAN}), E230H/2263, and E230H/2713 (see \href{https://www.stsci.edu/contents/news/stis-stans/july-2023-stan}{July 2023 STAN}). These changes however are not applicable to observations made pre-SM4. 

Prior to August 2002, echelle spectra underwent monthly offsetting in the spatial and spectral direction to more evenly distribute illumination across the detector. This monthly offsetting in the spatial and spectral directions in pre-SM4 observations greatly increases the difficulty of a full recalibration of each echelle mode. A flux calibration effort like the one described above for post-SM4 data is not possible for pre-SM4 observations with the data available in the archive. In this ISR, we describe an alternative strategy for the recalibration of pre-SM4 era echelle observations. One primary change in the CALSPECv11 library is the increase in the continuum flux of G191-B2B by 1-3\%. We measure the ratio of this new continuum to the model used to calibrate pre-SM4 observations, CALSPECv04. We then apply this ratio as a wavelength-dependent scaling to the previously derived sensitivities of each echelle mode (\href{https://www.stsci.edu/files/live/sites/www/files/home/hst/instrumentation/stis/documentation/instrument-science-reports/_documents/200701.pdf}{Aloisi et al. 2007}), and do not change the previously-derived blaze-shift coefficients. This results in an overall increase in flux so that pre-SM4 observations are more consistent with CALSPECv11. We also compute a wavelength-dependent ratio (CALSPECv11/CALSPECv07) that can be used to recalibrate post-SM4 observations in a similar manner. In the future, this simpler throughput scaling approach described in this ISR could instead be applied to post-SM4 observations with lower-priority echelle modes, requiring significantly less total effort. We emphasize that while the recalibration method described in Hernandez et al. (in prep.) is ideal for post-SM4 observations, it is significantly more labor-intensive.  

In Section \ref{sec:obs}, we list the observations that were used in the testing of the throughput scaling method described above. In Section \ref{sec:method}, we describe our method for spline fitting of CALSPEC model continua, In Section \ref{sec:res}, we show how these updates improve the flux calibration of pre-SM4 observations for a variety of STIS echelle modes, and we compare to the results of the post-SM4 flux recalibration strategy. We conclude with recommendations in Section \ref{sec:conc}.

%!!!!!!!!!!!!!!!!!!!!!!!!!!!!!!!!!!!!!!!!!!!!!!!!!!!!!!!!!!!!!!!!!!!!!!!!!!!!!!!!!!!!!!!!!!!!!!!!!!!!!!!!!!!!!!!!!!!!!!!!!!!!!!!!!!!!!!!!!!!!!!!!!!!!!!!!!!!!!!!!!!!!!!!!!!!!!!!!!!!!!!!!!!!!!!!!!!!!!!!!!!!!!!!!!!!!!!!!!!!!!!!!!!!!!!!!!!!!!!!!
%THIS SECTION NEEDS TO GO AFTER THE END OF THE FIRST PAGE AND BEFORE THE END OF THE SECOND PAGE
%Fill in Instrument, Year, and ISR Number and delete "newpage" immediately after this message
%!!!!!!!!!!!!!!!!!!!!!!!!!!!!!!!!!!!!!!!!!!!!!!!!!!!!!!!!!!!!!!!!!!!!!!!!!!!!!!!!!!!!!!!!!!!!!!!!!!!!!!!!!!!!!!!!!!!!!!!!!!!!!!!!!!!!!!!!!!!!!!!!!!!!!!!!!!!!!!!!!!!!!!!!!!!!!!!!!!!!!!!!!!!!!!!!!!!!!!!!!!!!!!!!!!!!!!!!!!!!!!!!!!!!!!!!!!!!!!
% \newpage

\lhead{}
\rhead{}
\cfoot{\rm {\hspace{-1.9cm} Instrument Science Report STIS 2024-02 Page \thepage}}
%%%%%%%%%%%%%%%%
\vspace{-0.3cm}
\ssection{Observations}\label{sec:obs}
The sensitivity scaling approach described in the section above was extensively tested on pre-SM4 observations of different echelle settings. In Table \ref{tab:obs} we list the exposure information of the echelle data used in our testing. Throughput updates were tested on pre-SM4 observations of the standard white dwarf G191-B2B.

\begin{table}[!h] % [!htb] forces it to be "here" "top" "bottom" and ! overwrites the LaTeX default
  \centering
    \caption{Summary of observations used to test scaling of pre-SM4 throughputs.}
    \def\arraystretch{1.25}
    \begin{tabular}{c c c c c c}
    \hline
    \hline
    Grating & Central Wavelength& Dataset & Observation Date & Exposure Time (s) \\
    \hline
    E140H & 1271 & o6hb10040 & 2001-09-17 & 640\\
    E140H & 1416 & o57u01020 & 1998-12-17 & 2040\\
    E140M & 1425 & o4dd16020 & 1999-02-06 & 4903\\
    E230H & 2263 & o6hb200d0 & 2001-09-18 & 609.6\\
    E230H & 2713 & o6hb30070 & 2001-09-19 & 900\\
    E230M & 1978 & o5i012010 & 2000-03-14 & 2304\\
    E230M & 2415 & o6hb40040 & 2001-09-12 & 244\\
    E230M & 2707 & o5i012020 & 2000-03-14 & 2880\\
    \hline
    \end{tabular}
    \label{tab:obs} 
\end{table}

%%% NEXT SECTION %%%
\vspace{-0.3cm}
\ssection{Methods}\label{sec:method}

In this section, we describe our procedure for determining the smooth wavelength-dependent ratios of arbitrary CALSPEC models. These ratios are then applied as a simple scaling to selected echelle mode throughputs in their PHOTTAB reference files. We have used CALSPEC models of the white dwarf standard star G191-B2B as the reference for all throughput updates. 

%% Subsection Example
\vspace{-0.3cm}
\ssubsection{Fitting the CALSPEC Model Continuum}\label{subsec:fit}

\begin{figure}
  \centering
  \includegraphics[width=2.8in]{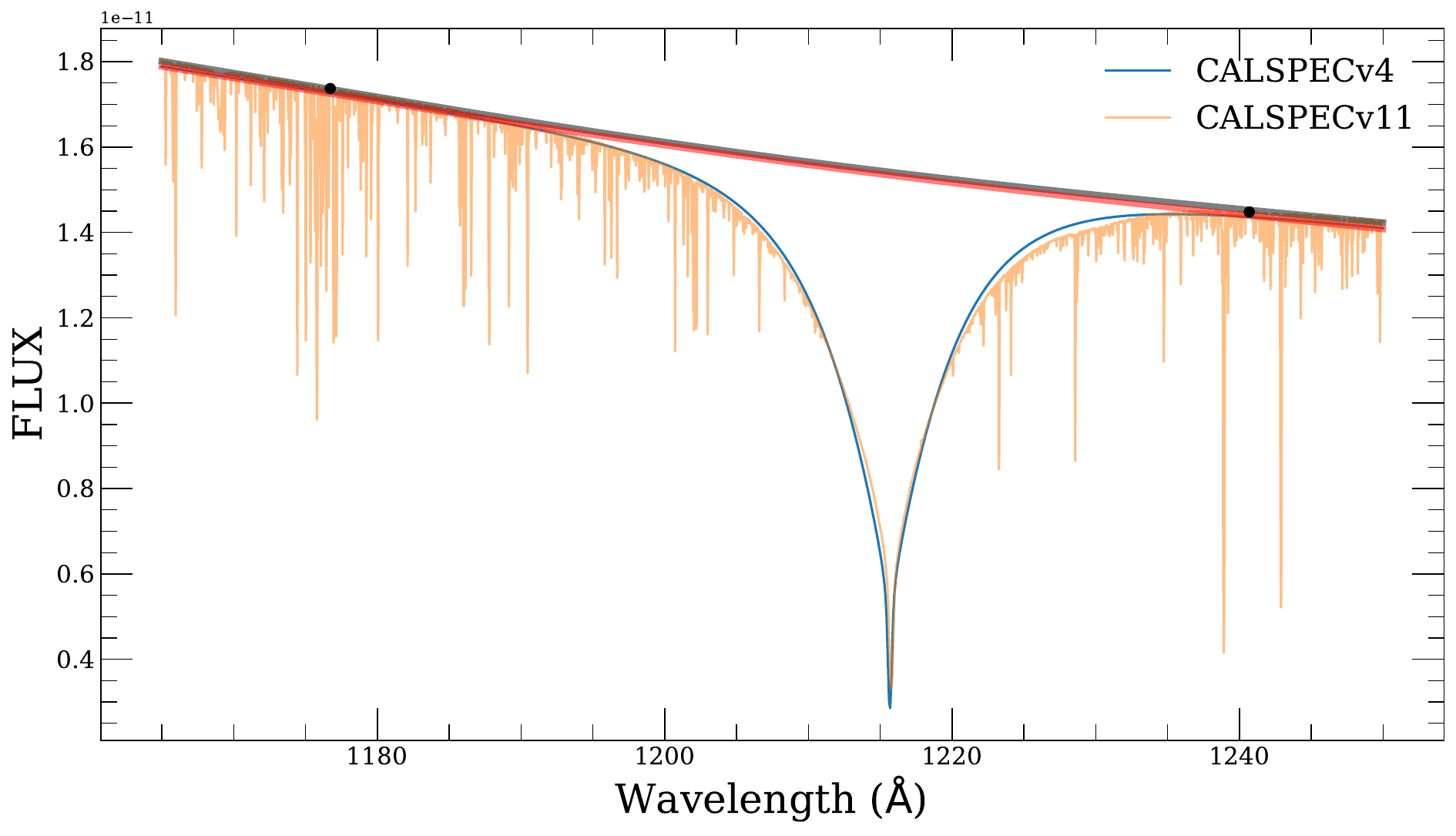}
  \includegraphics[width=2.8in]{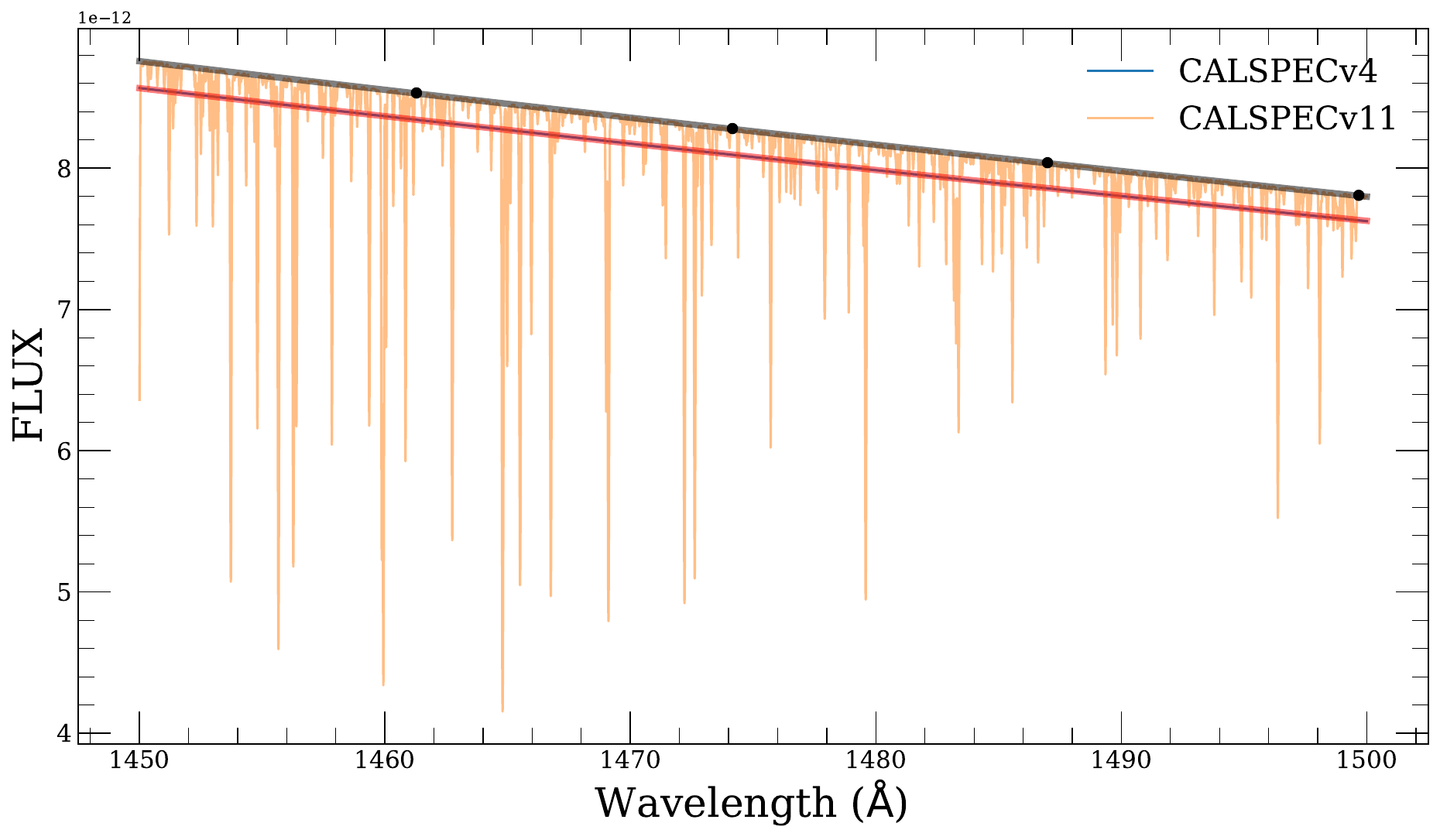}
  \includegraphics[width=2.8in]{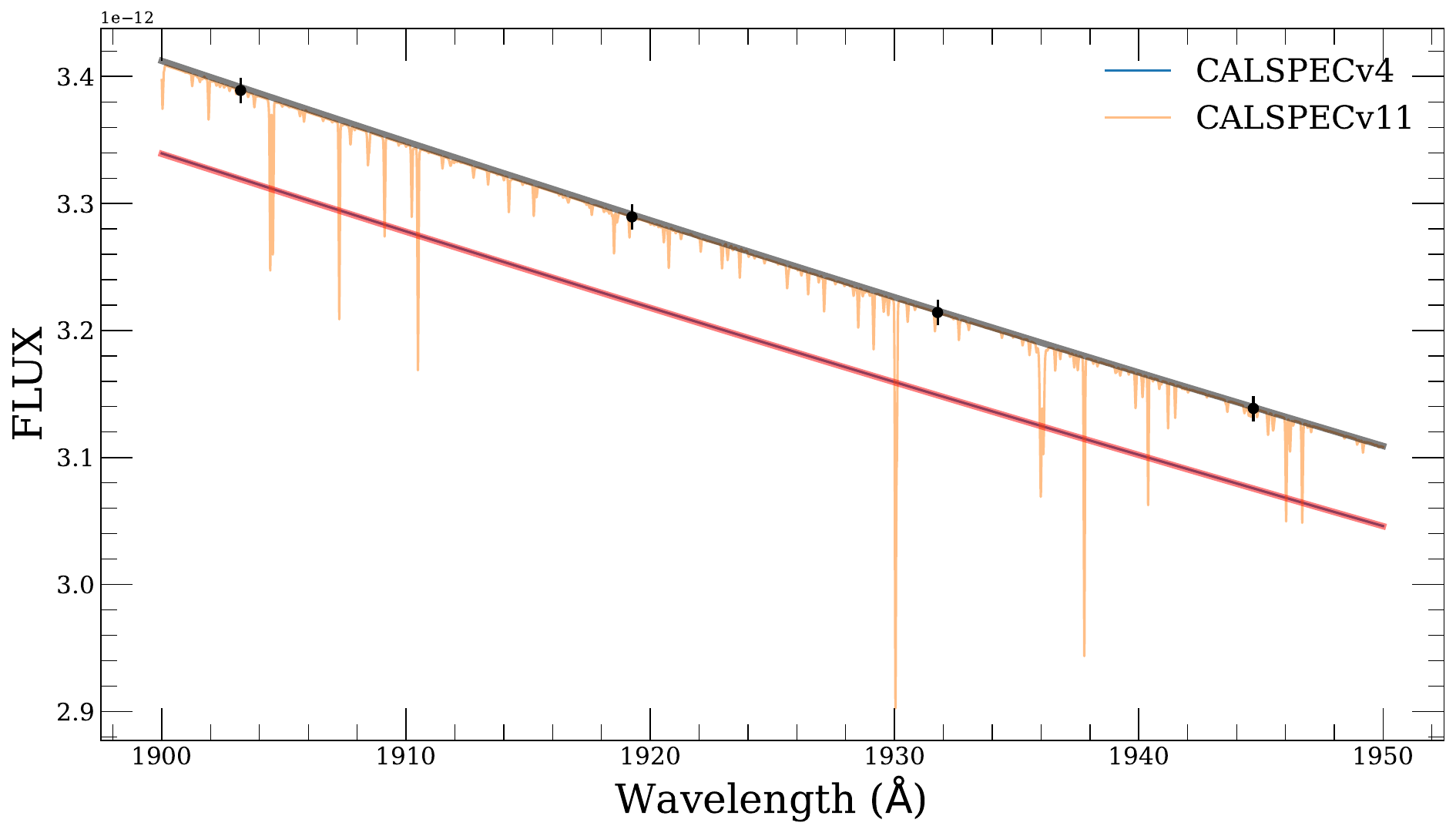}
  \includegraphics[width=2.8in]{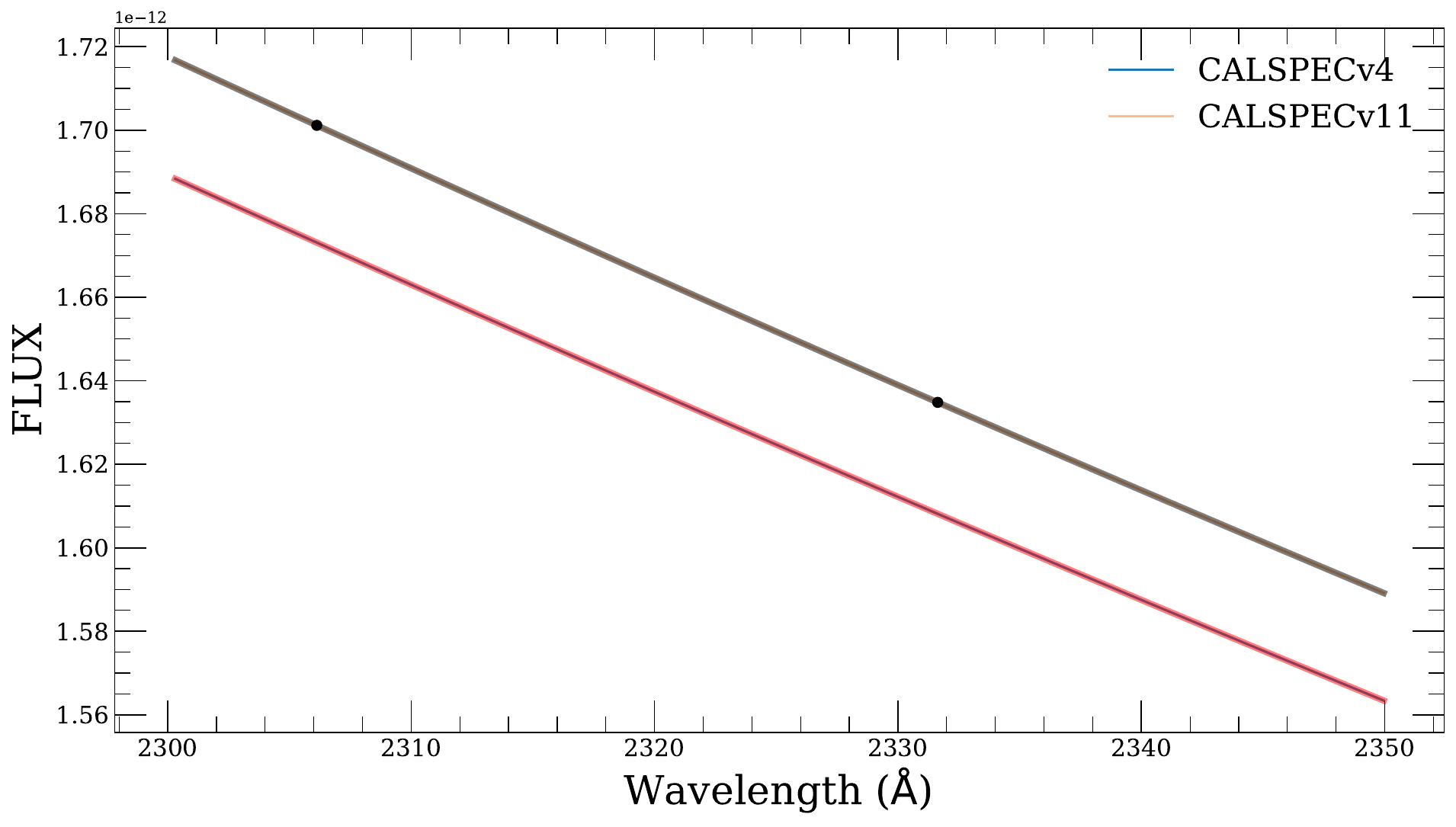}
    \caption{Comparison of CALSPECv04 (blue) and CALSPECv11 (orange) model spectra of G191-B2B and our continuum fits (red and black, respectively). Each panel covers a different wavelength range which in combination displays the representative differences in the model continua and absorption features. Below 2000\AA, CALSPECv11 includes a forest of narrow absorption features that we interpolate over with our spline fit. The black circles show the locations of the spline points that define the fit to the CALSPECv11 model of G191-B2B. Above 2000\AA, the continua of both models are smooth and well-represented by a spline fit with 50\AA-spacing between knots. In regions where the CALSPEC models are smooth, their curves lie directly under their respective spline fits.}
    \label{fig:fit}
\end{figure}

One significant change in the CALSPECv11 library is an updated metal-line-blanketed model atmosphere for G191-B2B. As shown in the panels in \autoref{fig:fit}, this results in a forest of narrow UV absorption features blueward of 2000\AA. Since we do not want to imprint these features on our wavelength-dependent throughput scaling, we have developed a procedure to automatically identify continuum points to be used in a spline fit. 

First, we smooth the CALSPEC model flux with an 11-pixel wide boxcar filter. We then compute the fractional deviation of the model flux from this smoothed flux as a function of wavelength. Based on this choice of smoothing, we have empirically determined that fluxes with a fractional deviation of $<0.001\%$ are likely to be continuum points that are unaffected by narrow absorption features. From these points, we identify all continuum sections where adjacent fluxes do not deviate by the same threshold. The mean wavelength and flux in each of these continuum sections is then identified as a spline knot location. This mediates the impact of defining spline knots near the edges of narrow absorption features. We require that adjacent spline knots have spacing $>12.5$\AA\ in order to ensure smooth monotonic variation of the spline fit to the continuum. We also avoid placing spline knots within the strong Lyman-$\alpha$ absorption feature ($1180-1240$\AA). Redward of 2000\AA\ where the continua in all models lack any absorption features, we use a linear 50\AA-spaced set of spline knot locations. We have assigned uncertainties of $1*10^{-14}$ erg s$^{-1}$ cm$^{2}$ \AA$^{-1}$ and $1*10^{-18}$ erg s$^{-1}$ cm$^{2}$ \AA$^{-1}$, for spline knot blueward and redward of 2000\AA, respectively. A fourth order spline is fit to the specified continuum locations of both the CALSPECv11 and the relevant comparison model (CALSPECv04 for pre-SM4 observations) using identical positions for spline knots.

Sections of the resulting fits are shown in \autoref{fig:fit}. In each panel, the orange and blue curves are the CALSPECv11 and CALSPECv04 G191-B2B models, respectively. The black and red curves are the resulting spline fits. The black points show our CALSPECv11 spline locations. These wavelength sections are representative of the entire fit, covering the Lyman-$\alpha$ feature (top-left), regions with strong narrow line blanketing (top-right and bottom-left), and regions where the model continua are smooth (bottom-right). Spline knots have been automatically selected in a systematic way that results in little impact on the continuum from adjacent narrow absorption features. 

\begin{figure}
  \centering
  \includegraphics[width=5.5in]{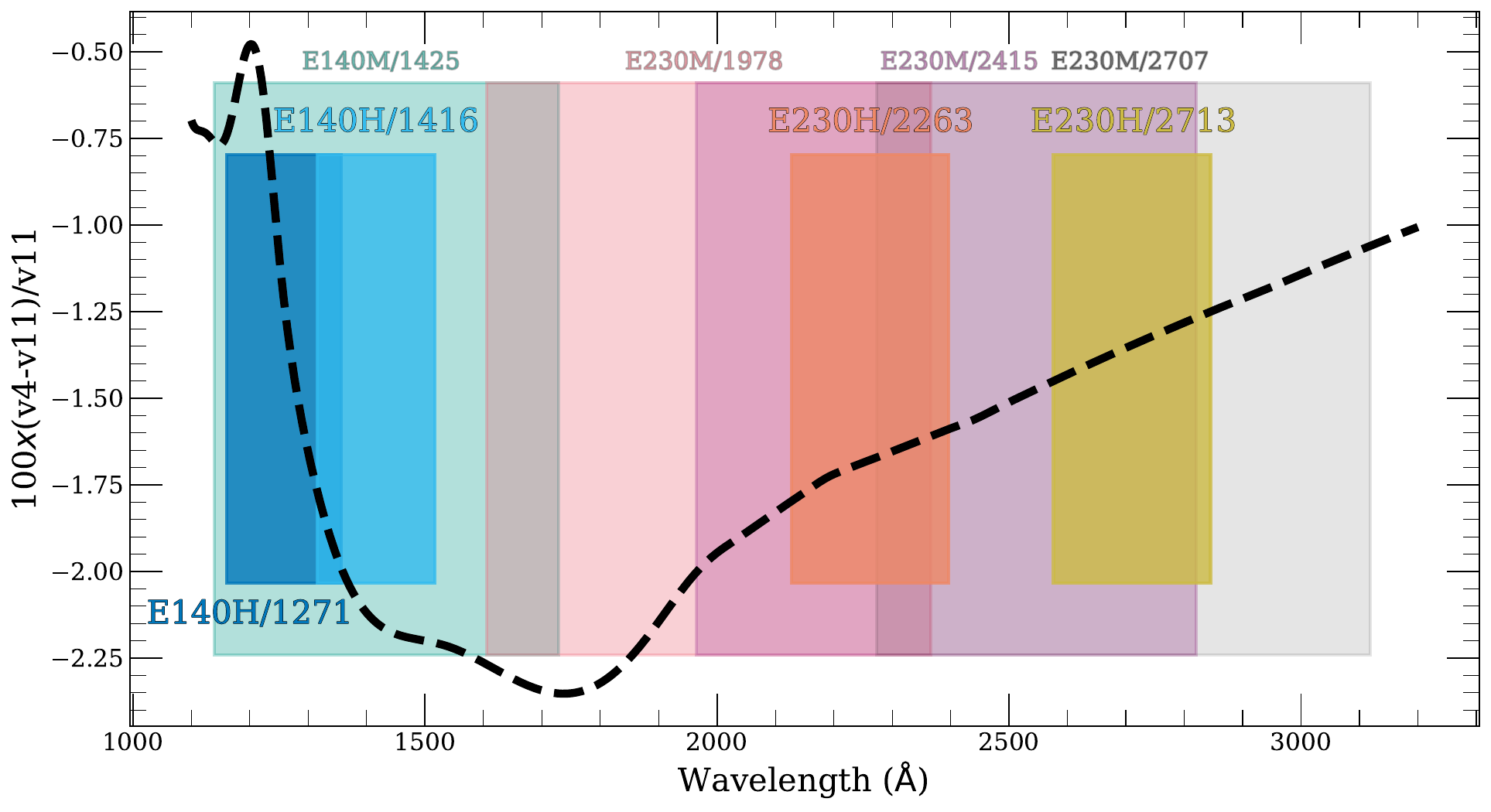}
    \caption{Percentage difference between spline fits to the CALSPECv04 and CALSPECv11 G191-B2B models (black dashed curve).  Colored shaded regions show the wavelength ranges of echelle modes for which we have updated pre-SM4 throughputs. Contintuum differences range from 0.48 - 2.40\%.}
    \label{fig:ratio}
\end{figure}

The scaling that is applied to the pre-SM4 throughputs is determined from the ratio of the black and red curves. We show this ratio as a function of wavelength expressed as a percentage of the CALSPECv11 model flux in \autoref{fig:ratio} (black dashed curve). The colored bands represent wavelength ranges of the echelle observing modes for which we have updated pre-SM4 throughputs using this ratio as a wavelength-dependent scaling factor. Across all of these modes, typical scaling values result in changes of 1-2\%. The largest difference is 2.4\% occurring at a wavelength of 1738\AA.

%% Subsubsection Example
\vspace{-0.3cm}
\ssubsection{Throughput scaling}\label{subsec:tput}

\begin{figure}
  \centering
  \includegraphics[width=1.81in]{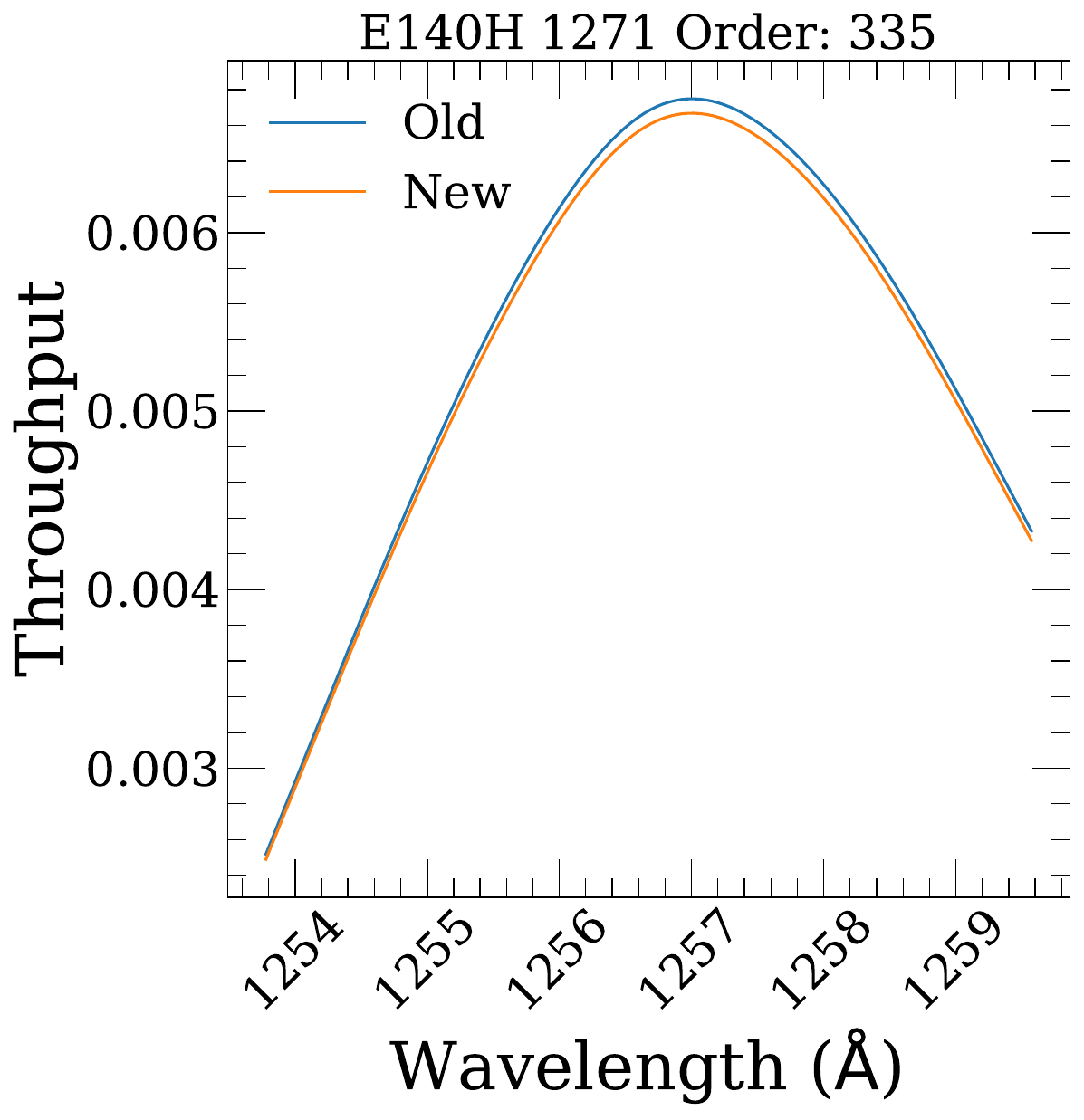}
  \includegraphics[width=1.88in]{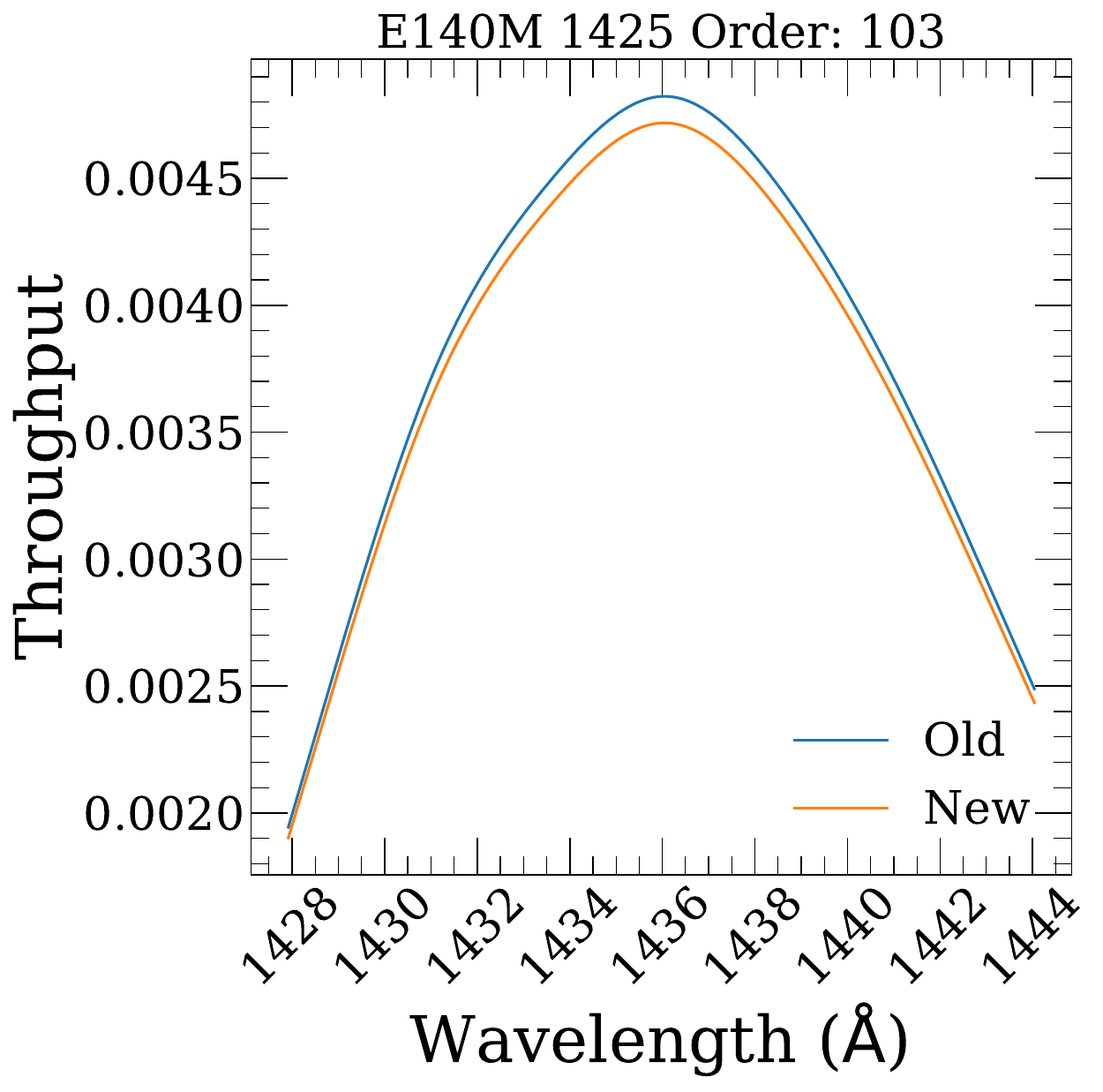}
  \includegraphics[width=1.85in]{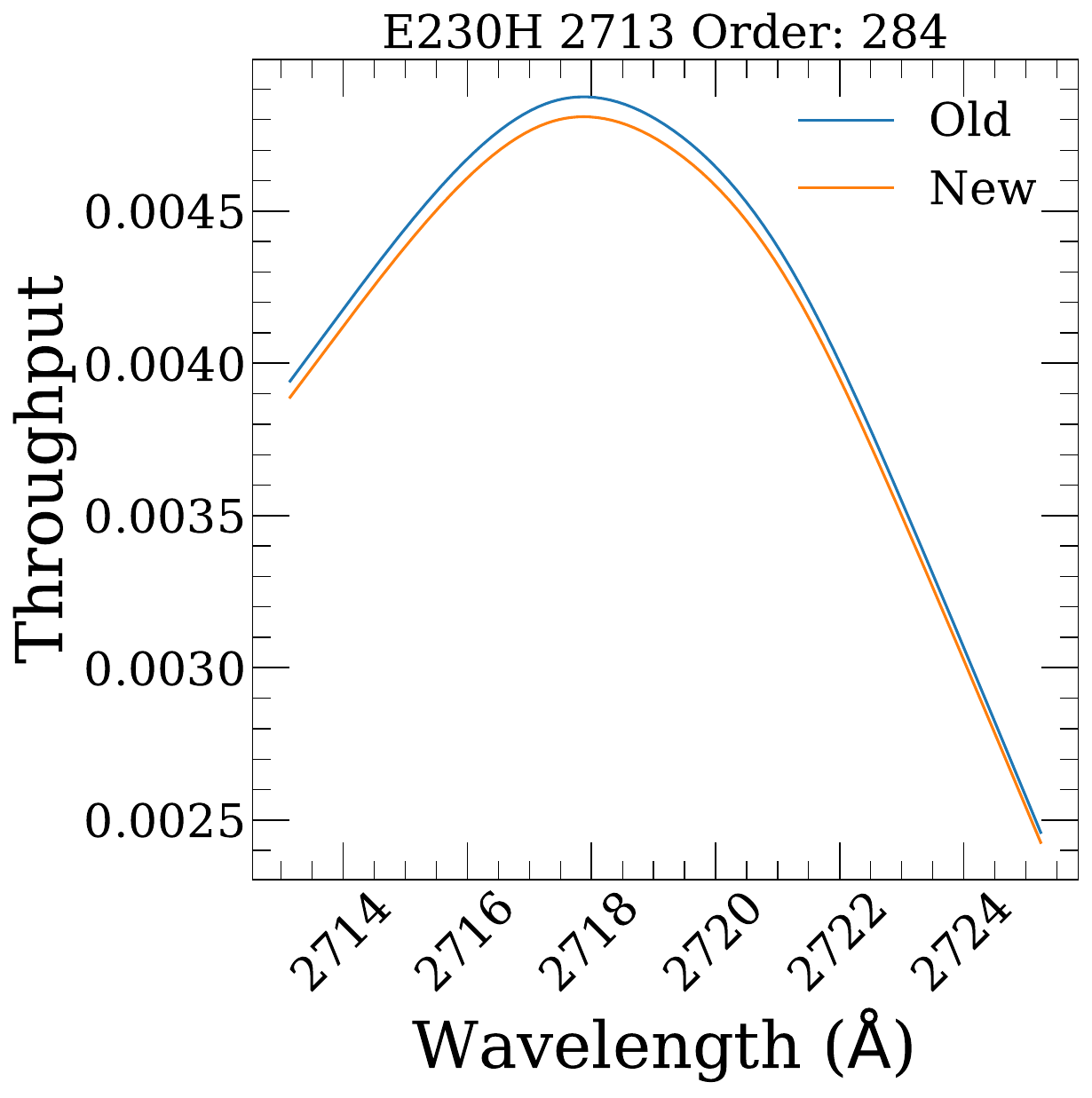}
    \caption{Example throughput updates of orders from E140H/1271, E140M/1425, and E230H/2713. Old and new throughputs are represented as blue and orange curves, respectively. New throughputs are lower to account for the overall increase in model continuum flux from CALSPECv04 to CALSPECv11. These orders cover wavelengths regions corresponding to the minimum, maximum, and typical flux differences ($\sim$ 0.5\%, 2.4\%, and 1.5\%, respectively).}
    \label{fig:tput}
\end{figure}

The improvements in the recent CALSPECv11 models are propagated to the echelle throughputs, which are defined as the efficiencies for an infinite aperture and infinite extraction box. These throughputs are stored in the STIS PHOTTAB reference files and are used by the CALSTIS pipeline. Here the increase in the overall continuum level from CALSPECv04 to CALSPECv11, is represented by an equivalent decrease in the effective throughput in each order. For each order of each grating, a wavelength-dependent scaling is applied based on the ratio of CALSPECv04 to CALSPECv11, determined from our spline fits. In \autoref{fig:tput}, we show example throughput updates to selected orders from E140H, E140M, and E230H (left to right), covering most of the FUV to NUV wavelength range.

%% ANOTHER SECTION
\vspace{-0.3cm}
\ssection{Results and Discussion}\label{sec:res}

\begin{figure}[htb!]
  \centering
  \includegraphics[width=5.7in]{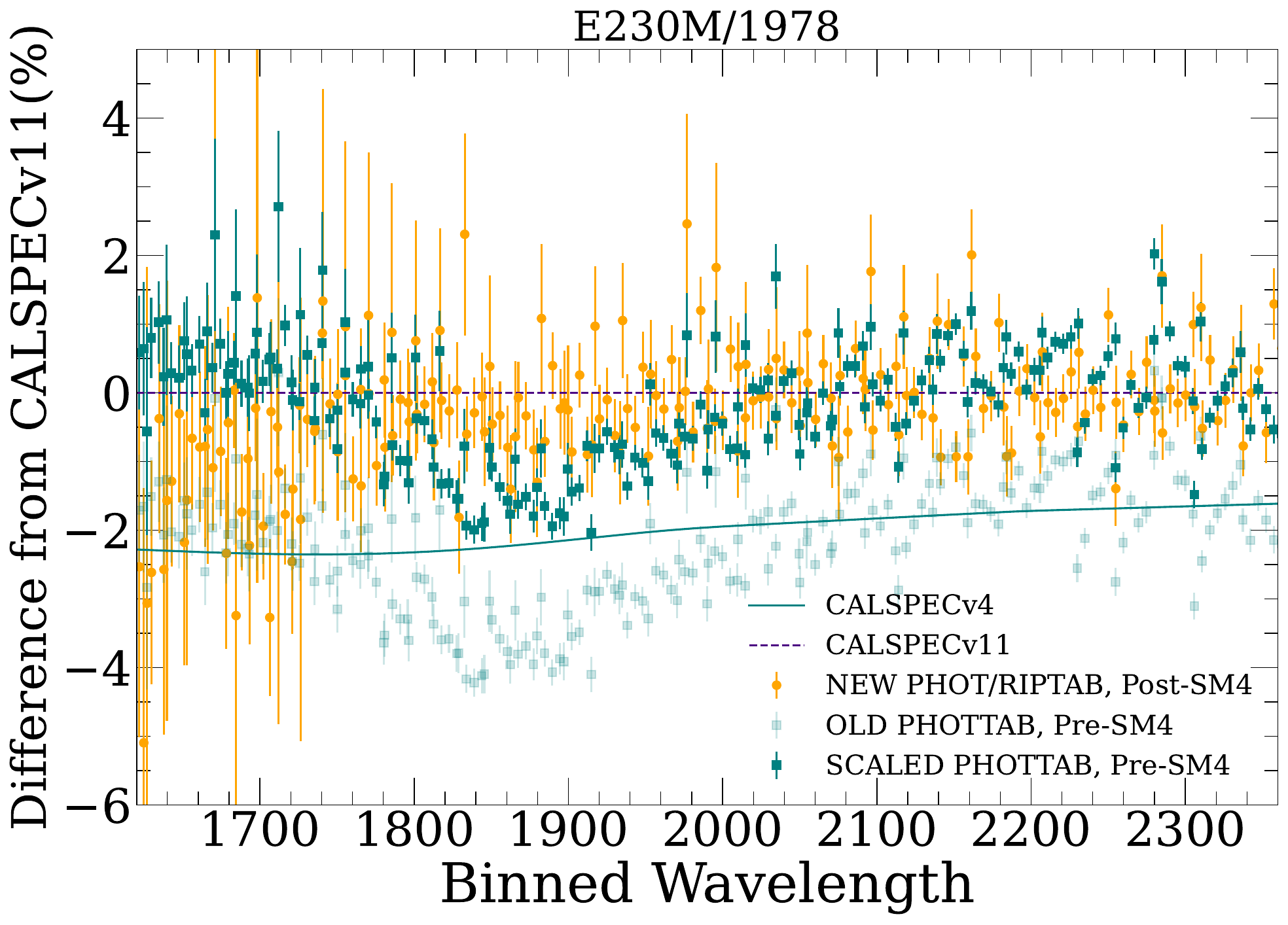}
    \caption{Example flux calibration improvement from scaling of the pre-SM4 throughput of E230M/1978 using the average difference from CALSPECv11 as a function of wavelength. Light green points are from the original unscaled throughput, dark green points are using the new scaled throughput based on the curve shown in \autoref{fig:ratio}, and the dark green curve is CALSPECv04. The orange points are the results of the new flux recalibration of this mode for post-SM4 observations (Hernandez et al. in prep). The median difference from CALSPECv11 using the old, scaled, and new phottab, is -2.06, -0.10, and -0.22$\%$, respectively. The simple scaling approach for pre-SM4 data results in a significant improvement in flux calibration, however, it cannot account for changes in the shape of the sensitivity function.}
    \label{fig:e230m_res}
    \vspace{5mm}
\end{figure}

In this section, we investigate the improvements that are possible using this pre-SM4 throughput scaling approach to flux recalibration. In January of 2023, the STIS team updated the flux calibration of the primary echelle mode E230M/1978 for post-SM4 observations (see \href{https://www.stsci.edu/contents/news/stis-stans/january-2023-stan#article1}{Jan 2023 STAN}). This update included revised throughputs that were calibrated directly to the CALSPECv11 library of models. In \autoref{fig:e230m_res}, we compare the results of that direct recalibration to the results from our scaling approach in post-SM4 data to results using our scaling approach in pre-SM4 data. 

This figure shows the average difference in flux from the G191-B2B CALSPECv11 model and the calibrated exposures using the different updated throughputs expressed as a percentage as a function of wavelength with 5\AA\ binning. The light green squares are the results of running calstis with the original pre-SM4 PHOTTAB files (throughputs derived based on CALSPECv04 models, with no additional scaling) and the dark green squares are the results after applying throughput scaling. The orange points instead use the newly derived PHOTTAB and RIPTAB files on post-SM4 observations of the same standard star. It is clear that throughput scaling does on average improve the flux calibration of these data by about $2\%$. We note that while the post-SM4 flux differences have less scatter and broad structure, the pre-SM4 scaling approach significantly improves the agreement with the CALSPECv11 models. At the time of writing, pre-SM4 throughput updates have thus far only been applied to the eight highest priority echelle modes, based on historical usage. Similar improvements are observed across each of these modes and we have included similar comparisons in \autoref{fig:all_modes} in Appendix A. A summary of these improvements are shown in Table \ref{tab:improve}. 

\begin{table}[!h]
    \centering
    \caption{Summary of improvements seen in pre-SM4 echelle observations using the throughput scaling approach.}
    \begin{adjustbox}{center}
    \def\arraystretch{1.25}
    \begin{tabular}{c c c c c}
    \hline
    \hline
    Grating/ & Scaling &  Median Difference & Median Difference & Median Difference\\
    Central & Range ($\%$) & from CALSPECv11 &from CALSPECv11 & from CALSPECv11\\
    Wavelength  &  & Before Scaling ($\%$) & After Scaling ($\%$) & Post-SM4 ($\%$) \\
    \hline
    E140H/1271$^*$  & 0.7-2.0 & -2.11 & -0.55 & -\\
    E140H/1416  & 1.8-2.3 & -2.77 & -0.68 & -\\
    E140M/1425$^*$  & 0.8-2.4 & -2.37 & -0.31 & -0.35\\
    E230H/2263  & 1.8-1.6 & -1.72 & -0.04 & -0.12\\
    E230H/2713  & 1.5-1.3 & -2.01 & -0.69 & -0.03\\
    E230M/1978  & 2.3-1.7 & -2.06 & -0.10 & -0.22\\
    E230M/2415  & 2.0-1.3 & -1.98 & -0.35 & -0.13\\
    E230M/2707  & 1.7-1.1 & -1.65 & -0.34 & 0.11\\
    \hline
    \end{tabular}
    \label{tab:improve} 
    \end{adjustbox}
    \newline
    $^*$Median differences computed using wavelengths unaffected by Ly-$\alpha$ absorption.
\end{table}

\begin{figure}
  \centering
  \includegraphics[width=5.5in]{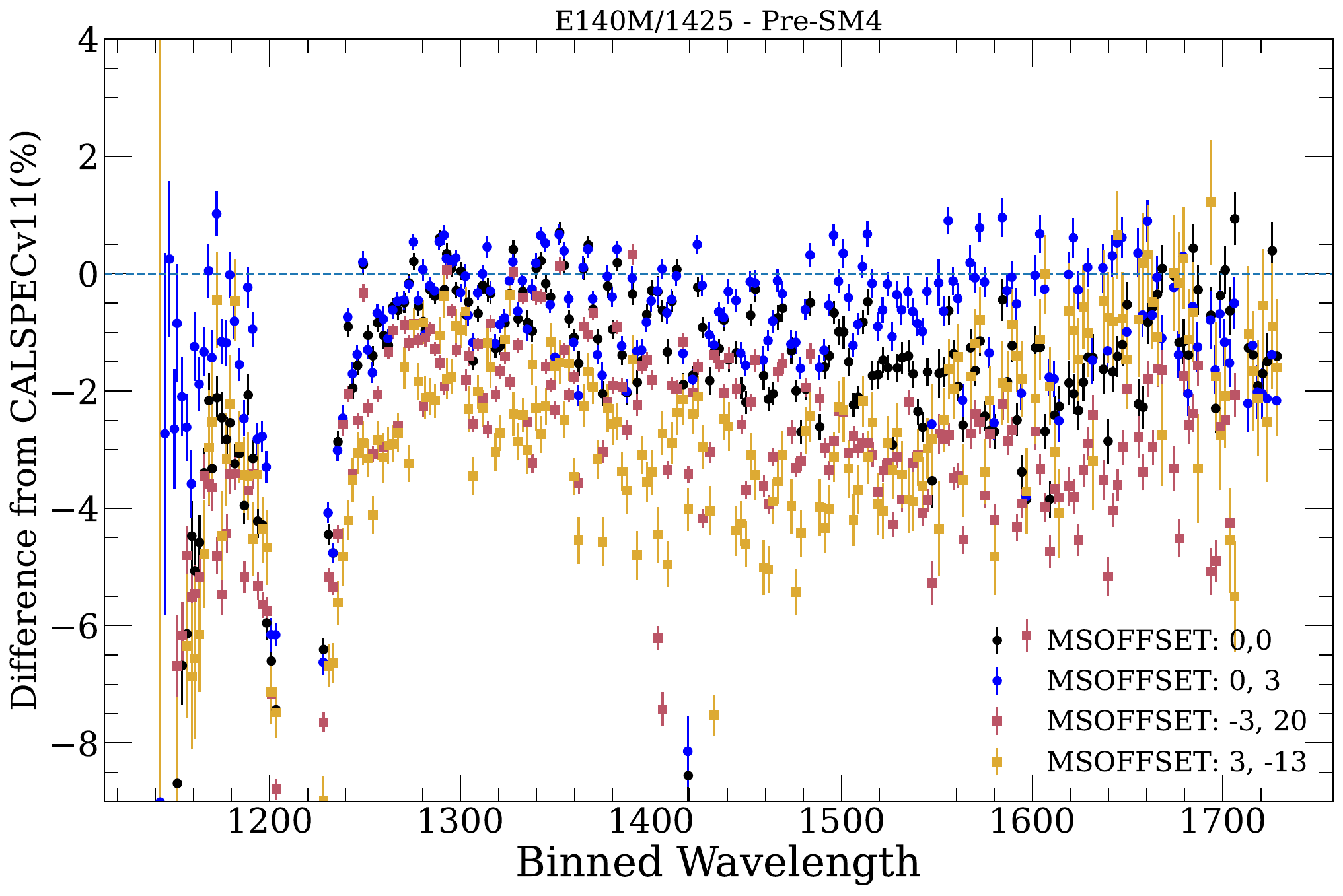}
    \caption{Similar to \autoref{fig:e230m_res} but for pre-SM4 observations with E140M/1425 with different monthly offset settings (sets of colored points). The median difference from CALSPECv11 for is -1.33, -0.54, -2.73, -2.48 $\%$ for the 0/0, 0/3, -3/20, and 3/-13 X/Y offset settings, respectively. The largest differences occur for observations using larger monthly offsets. However, since throughputs have been decreased by $0.8 - 2.4\%$ over this wavelength range (see Table \ref{tab:improve}), the difference from CALSPECv11 represents an improvement in all settings over the original reference files. }
    \label{fig:month}
\end{figure}

One additional concern with pre-SM4 observations is the monthly spatial and spectral offsetting of the STIS MAMA modes. Since observations taken with different monthly offsets illuminate different parts of the detector, it is reasonable to expect that this will impact flux calibration. We investigate the results of this method for a variety of monthly offsets in \autoref{fig:month}. Similar to \autoref{fig:e230m_res}, we show the average difference between the CALSPECv11 (G 191-B2B) model and echelle exposures calibrated with the scaled throughputs as a function of wavelength for different monthly offsets (sets of colored points). Overall, we find that the scatter in the data is larger for these different settings. For settings with the largest monthly offsets (red and yellow points), we find the largest deviation from CALSPECv11. However, we note that in all cases, fluxes at each wavelength have been scaled higher by $\sim 2\%$ and since none of the observations appear over-corrected we have improved the flux calibration of each observation. This trend is also representative of other echelle modes.

We caution that the number of standard star observations with different monthly offsets is very limited for each echelle mode. This prohibits a systematic study of our flux calibration improvements as a function of monthly offset, and limits our ability to identify trends. Therefore we emphasize that we expect the pre-SM4 throughput scaling approach to provide the greatest improvement when monthly offets are zero, and cannot guarantee improvement for every grating/cenwave setting. 

\vspace{-0.3cm}
\ssection{Conclusions}\label{sec:conc}

We have updated the pre-SM4 throughputs of 8 different STIS echelle modes using the ratio of the new CALSPECv11 model to the CALSPECv04 (previously used to derive the pre-SM4 throughputs) models of G191-B2B. These changes result in roughly 0.5\% to 2.4 \% improvements across FUV and NUV wavelength ranges for these data. We have inspected how this method is impacted by the presence of monthly offsetting in pre-SM4 data. We also provide a procedure for implementing a similar solution for post-SM4 observations. On average we see improvement in flux calibration, however given the lack of data, we cannot guarantee that this method provides an improvement for all monthly offset settings. This simple scaling approach provides a lower-effort option to a complete flux recalibration that may be preferred for lesser used echelle modes. 

%%% DELETE THESE SECTIONS IF YOU DON'T USE THEM %%%

% %%% ACKNOWLEDGEMENTS %%% 
% \vspace{-0.3cm}
% \ssectionstar{Acknowledgements}
% \vspace{-0.3cm}
% Write any acknowledgements you want to here.

%%% CHANGE HISTORY %%%
\vspace{-0.3cm}
%Put instrument, year, and ISR number
\ssectionstar{Change History for STIS ISR 2024-02}\label{sec:History}
\vspace{-0.3cm}
Version 1: 15 March 2024 - Original Document %Month DD, YYYY format

%%% REFERENCES %%%
\vspace{-0.3cm}
\ssectionstar{References}\label{sec:References}
\vspace{-0.3cm}

\noindent
Aloisi, A., Bohlin, R., \& Kim Quijano, J. 2007, STIS Instrument Science Report 2007- 01\\
Bohlin, R., Hubeny, I., \& Rauch, T. 2020, AJ, 160, 21\\
Bostroem, K. A., Aloisi, A., Bohlin, R., Hodge, P., \& Proffitt, C. 2012, STIS Instrument Science Report 2012-01\\
Carlberg, J., Monroe, T., Riley, A., \& Hernandez, S., 2022, STIS Instrument Science Report 2022-04\\
Hernandez et al. 2024 in prep, STIS Instrument Science Report

%%% Appendix %%%%
\newpage
\vspace{-0.3cm}
\ssectionstar{Appendix A}\label{sec:Appendix}
\vspace{-0.3cm}

\begin{figure}[htb!]
  \centering
  \includegraphics[width=2.5in]{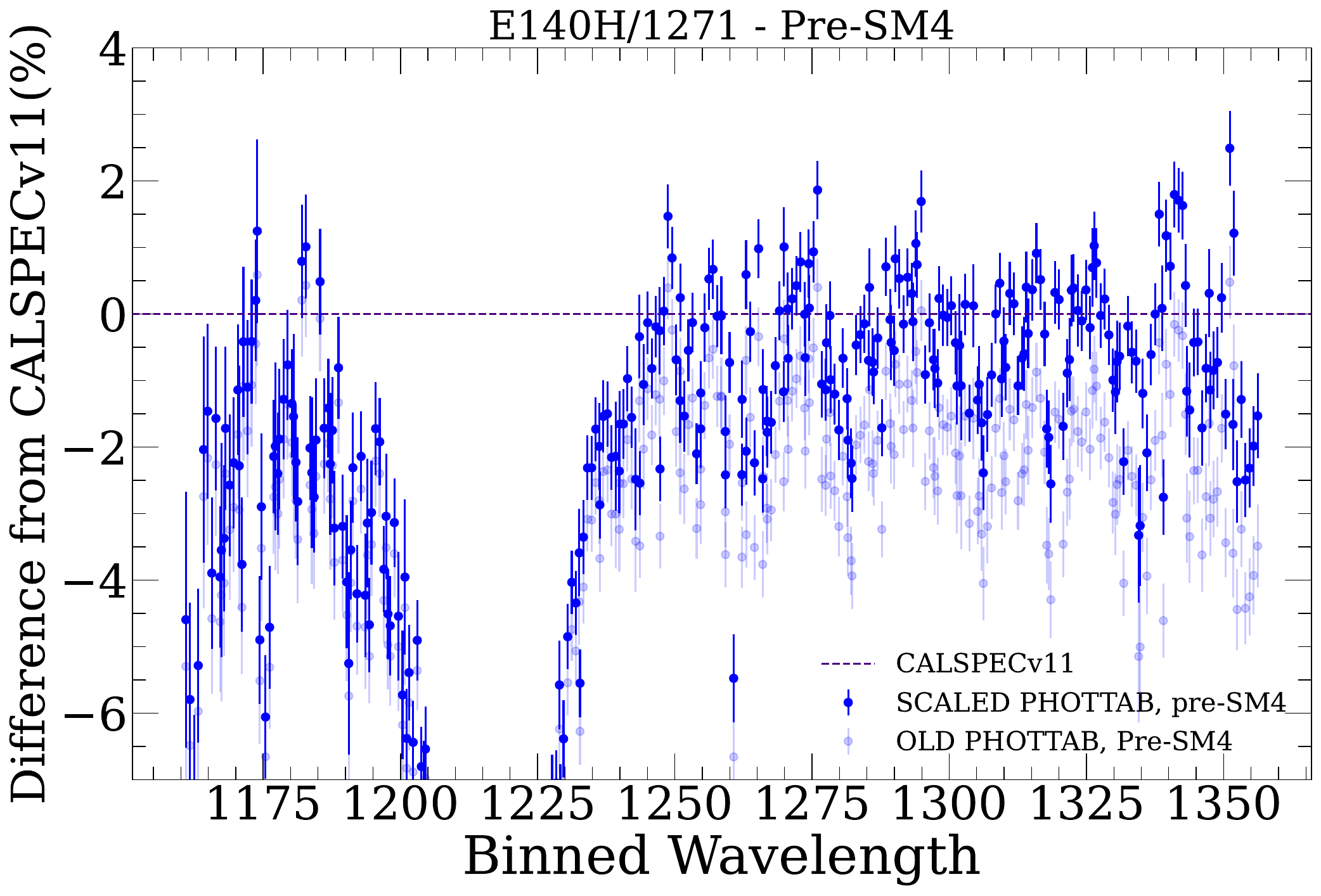}
  \includegraphics[width=2.5in]{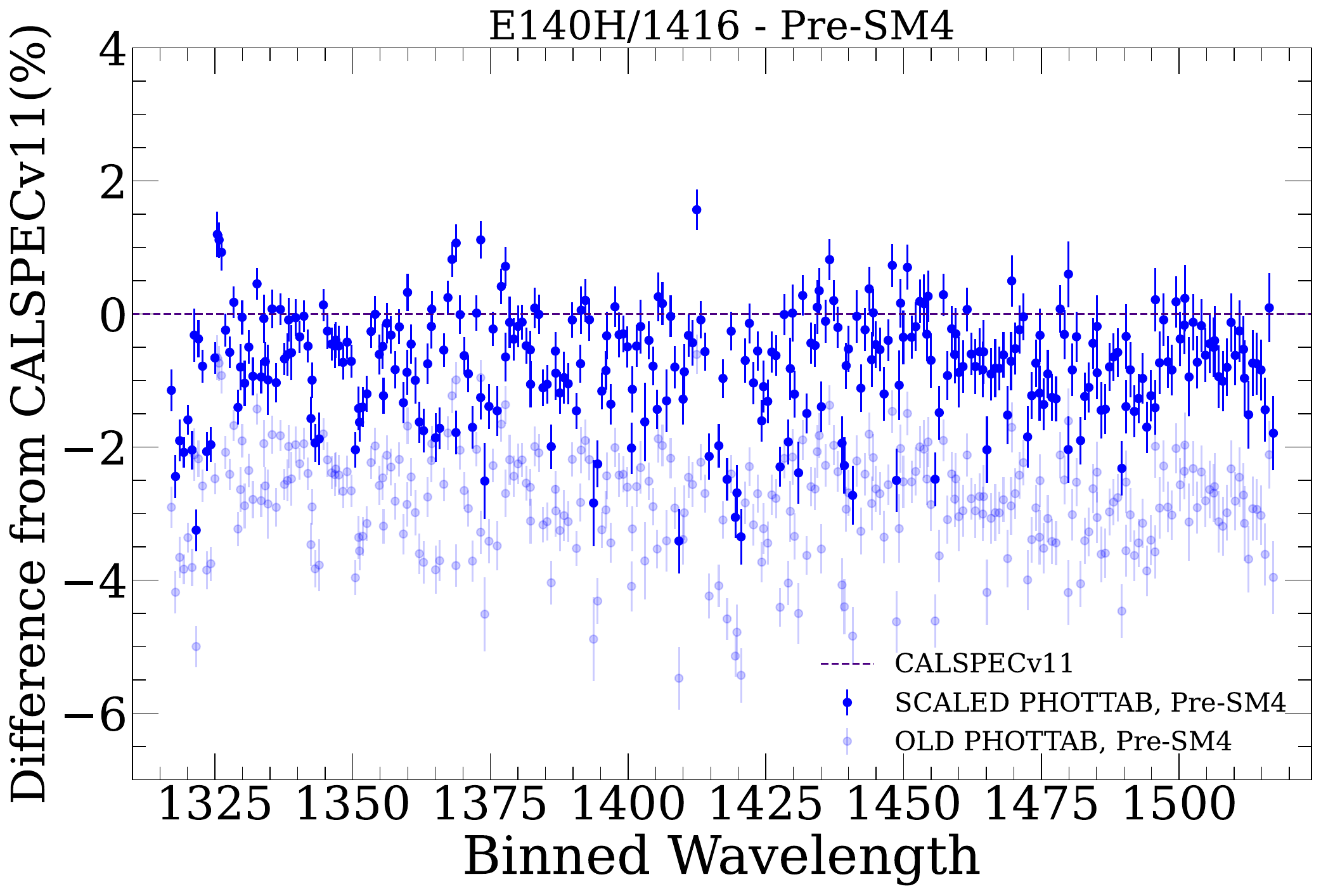}
  \includegraphics[width=2.5in]{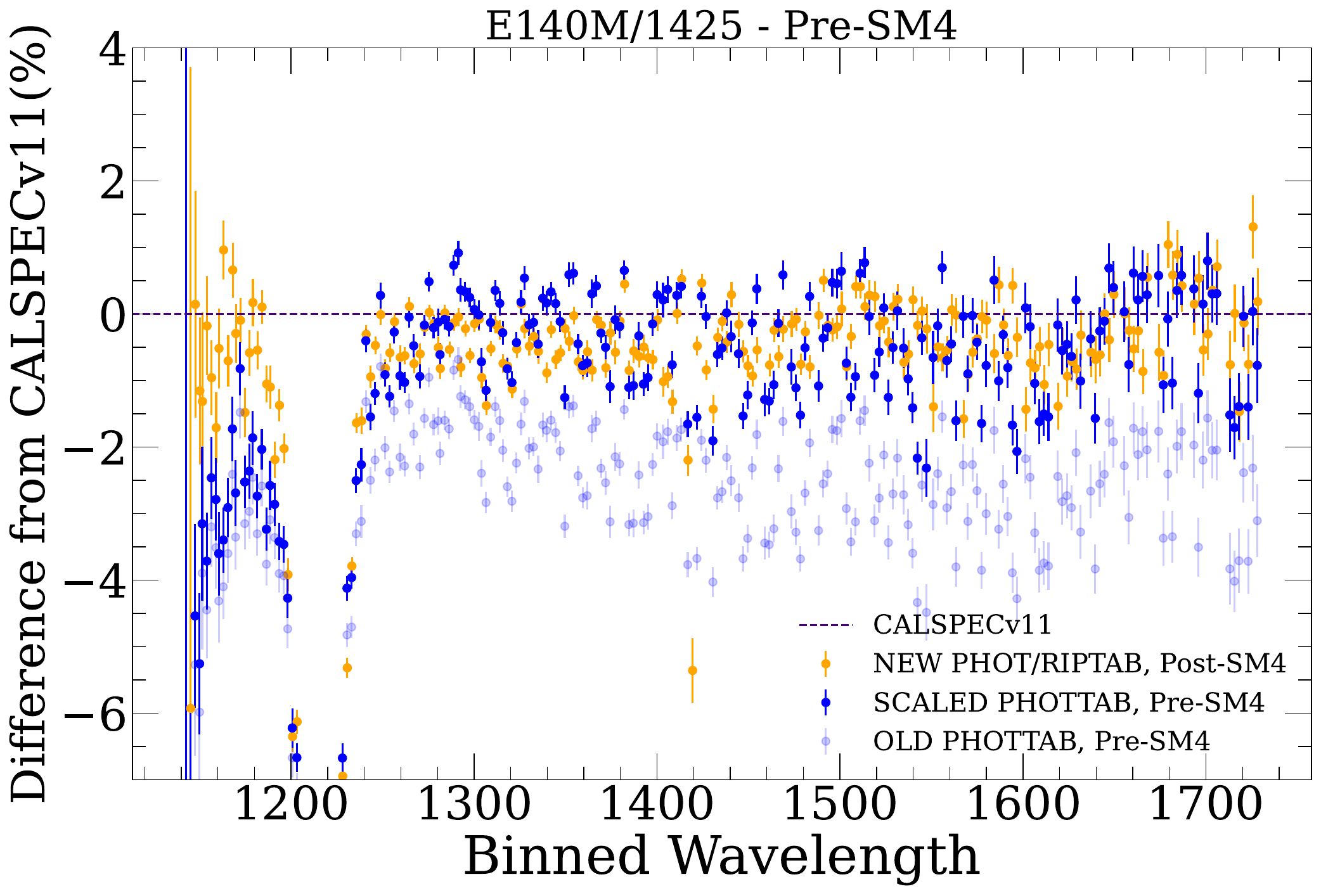}
  \includegraphics[width=2.5in]{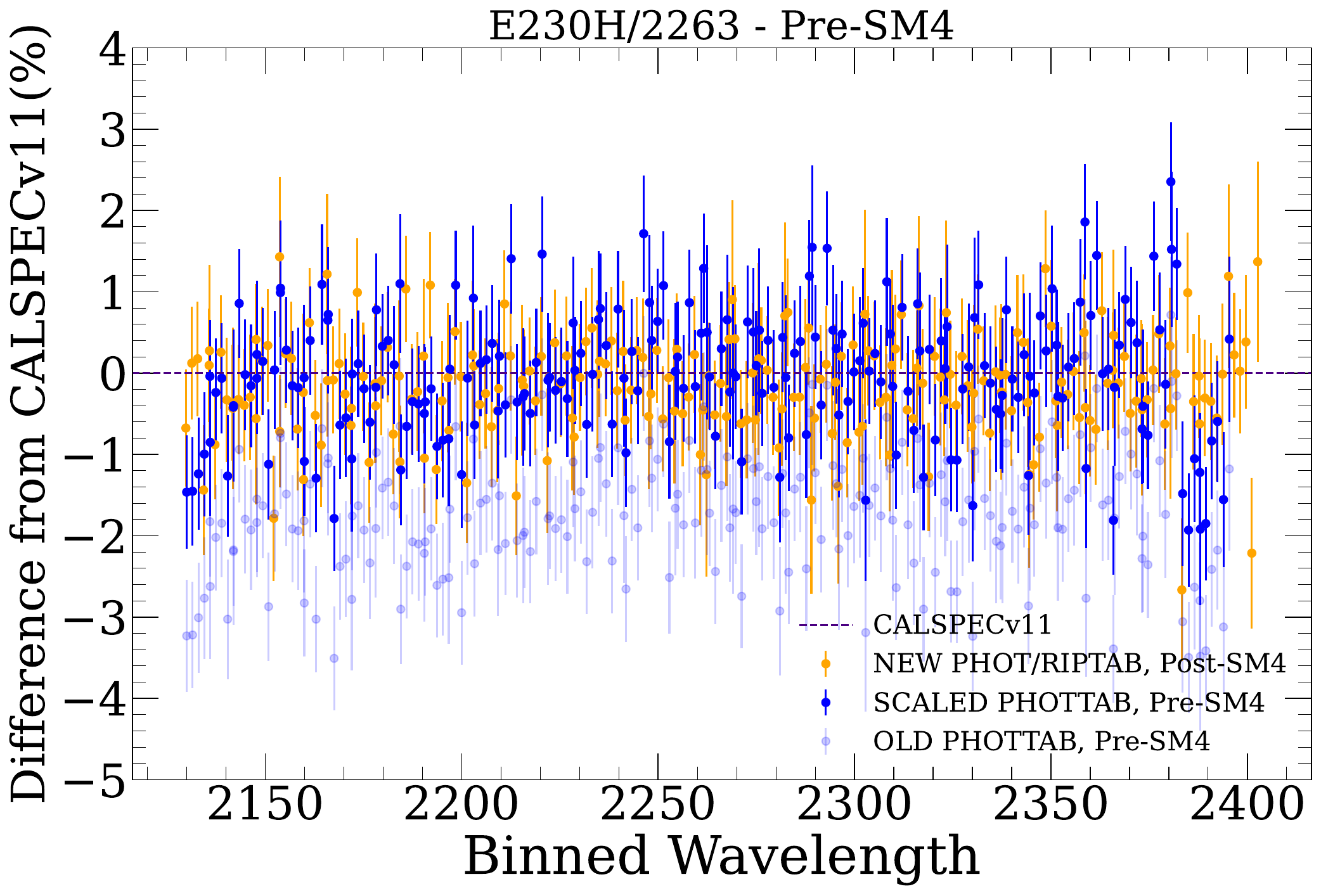}
  \includegraphics[width=2.5in]{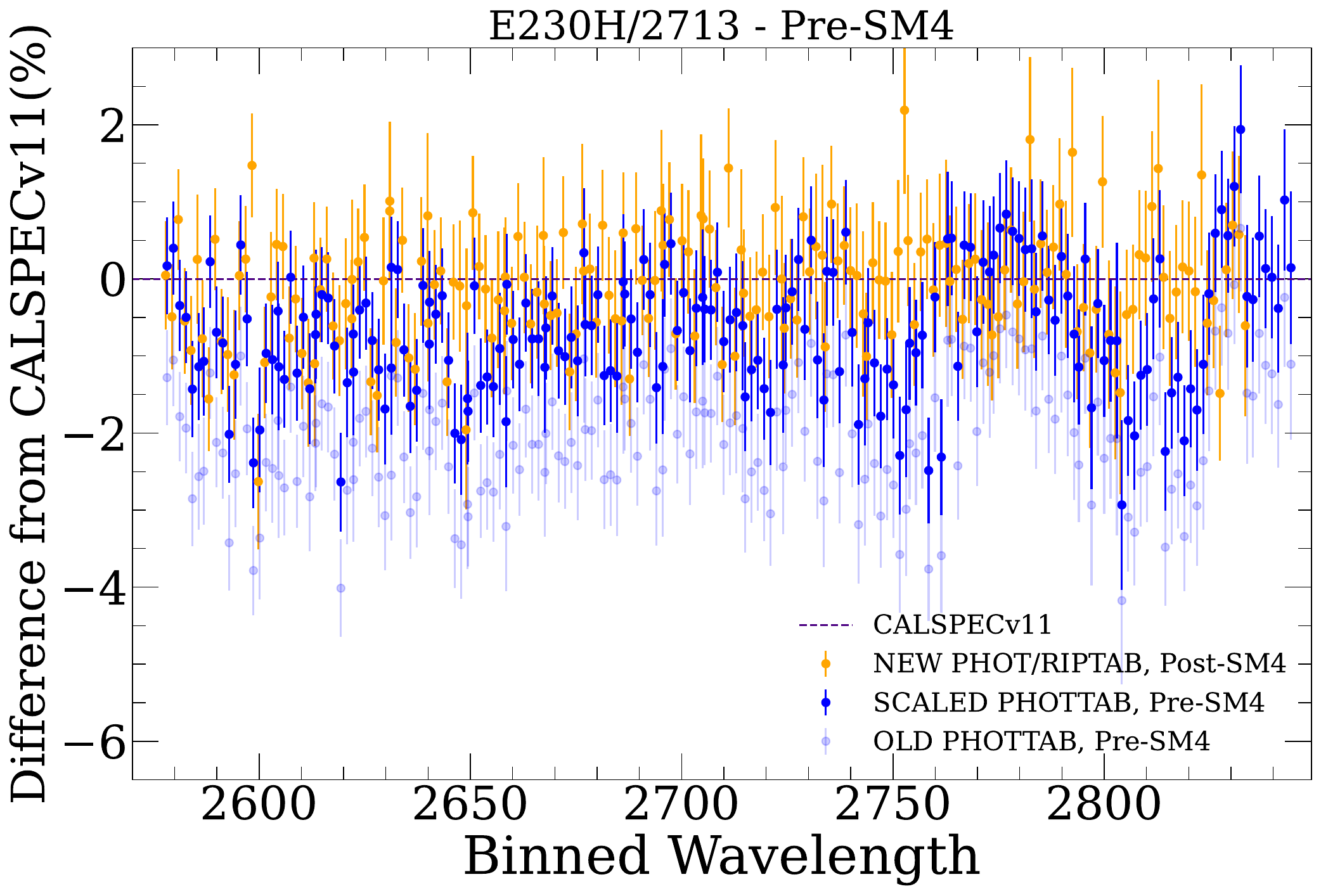}
  \includegraphics[width=2.5in]{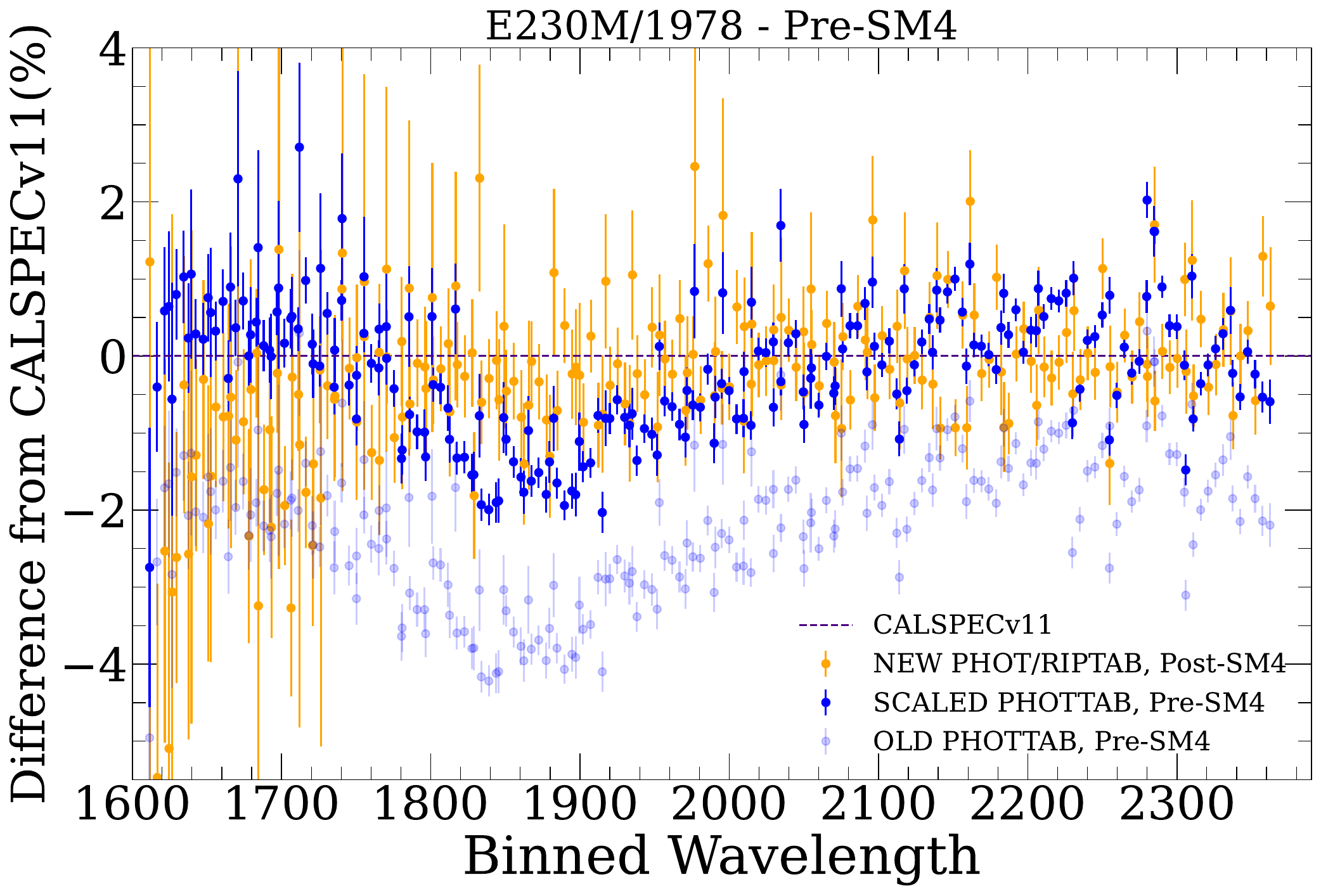}
  \includegraphics[width=2.5in]{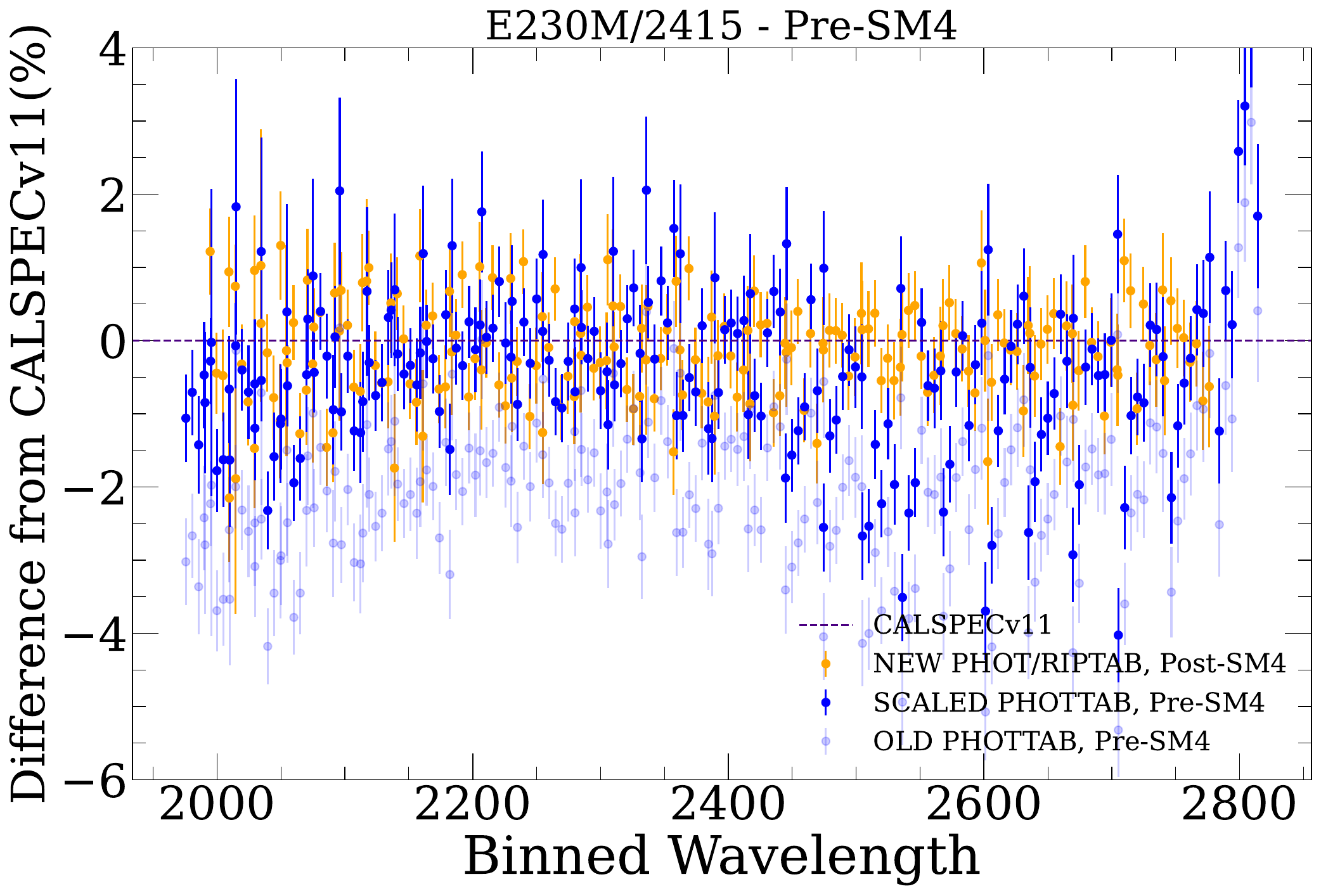}
  \includegraphics[width=2.5in]{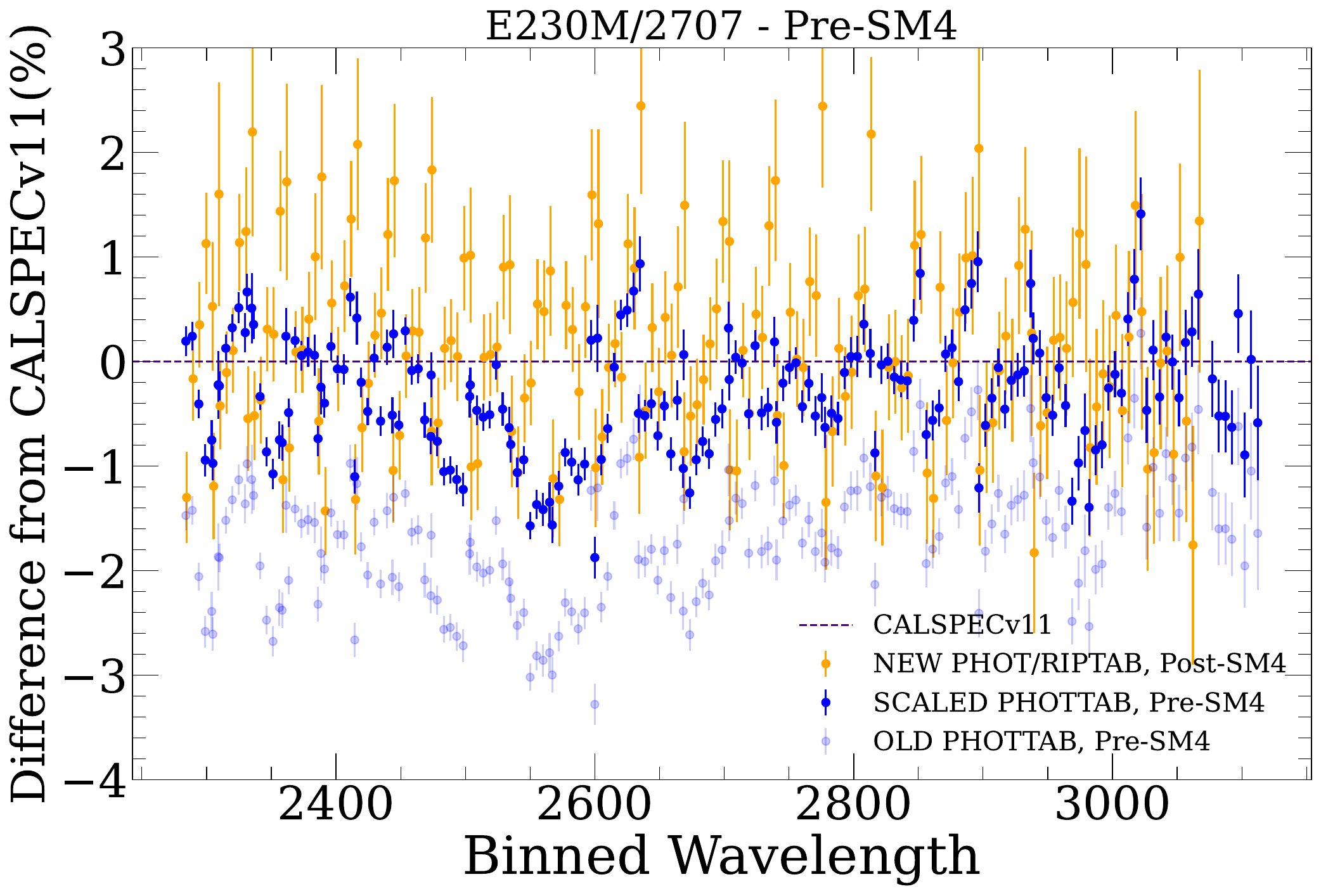}
    \caption{Example flux calibration improvements seen in each of the updated echelle modes characterized by the difference from CALSPECv11 as a function of wavelength. Panels are similar to Figure \ref{fig:e230m_res}. Light blue points are results using the original throughputs and dark blue points are results using the new scaled throughputs. Orange points show the post-SM4 observations of the same star calibrated with the newest reference files. }
    \label{fig:all_modes}
\end{figure}

\end{document}